\documentstyle[psfig]{mn}

\title[Near IR spectroscopy of ULIRGS]{Investigating the central engine and excitation mechanisms of ULIRGs: near infrared spectroscopy}
\author[A.J.Burston et al.]{A.J.~Burston,$^1$\thanks{email :
abu@star.le.ac.uk}  M.J.~Ward,$^1$ and  R.I.~Davies,$^2$\\ 
       $^1$X-ray Astronomy Group, University of Leicester, University
       Road, Leicester, LE1 7RH, UK.\\
       $^2$Max-Planck-Institut f\"{u}r extraterrestrische Physik,
       Garching, Germany.}

\def\iras{{\it IRAS}}

\begin{document}
\maketitle

\begin{abstract}
We present NIR observations of a sample of mainly interacting ULIRGs,
comprising H and K band spectra. Our main aims are to investigate the
power source of these extremely luminous objects and the various
excitation mechanisms of the strong molecular hydrogen emission often
seen in such objects. Broadened emission lines were only detected in
one object, IRAS\,23498, consistent with previous results for this
galaxy. The [SiVI] emission line was detected in IRAS\,17179 and
IRAS\,20210, both classified as Sy2s. Two of the sample were
unclassified, IRAS\,00150 and IRAS\,23420, which exhibit neither
[SiVI] emission or broadened HI emission. However this does not rule
out the presence of an obscured AGN. Analysis of the molecular
hydrogen emission showed that the major excitation mechanism for most
objects was thermal. Modelling of the more luminous objects indicate
for IRAS\,20210 10 per cent, and for both IRAS\,23365 and IRAS\,23420,
30 per cent of their 1--0S(1) line emission has a non-thermal origin.

\end{abstract}

\normalsize
\section{Introduction}
The existence of ultra-luminous infrared galaxies (ULIRGs) was
highlighted by the Infrared Astronomical Satellite (\iras). The class
is arbitrarily defined by Sanders and Mirabel \shortcite{sm96} as
galaxies having L$_{IR} \geq 10^{12}$ L$_\odot$, where L$_{IR} \equiv
$ L(8 -- 1000 $\mu$ m) using H$_{o}$ = 75 km s$^{-1}$ Mpc$^{-1}$ and
q$_{o}$ = 0. Note that, unless otherwise specified, we have adopted
H$_{o}$ = 50 km s$^{-1}$ Mpc$^{-1}$ and q$_{o}$ = 0 throughout. In
order to produce energy output of this magnitude, ULIRGs require a
substantial energy source. The two main candidates are a burst of
strong star formation in a central starburst or an active galactic
nucleus (AGN) with accretion onto a central black hole. ULIRGs are
generally very gas and dust rich e.g. Rigopoulou et
al.\shortcite{rig96}. Unfortunately this gas and dust obscures the
centre of the galaxies in which they reside, hampering identification
of the central source. Direct evidence for the presence of AGN is
found for some ULIRGs in the form of Seyfert-like emission line
ratios, high excitation lines and sometimes broad permitted-line
widths \cite{vs95}. It has been proposed that the fraction of ULIRGs
containing AGN rises with infrared luminosity (e.g. Sanders et al.
\shortcite{sd88}). Often evidence for the presence of strong star
formation is found (e.g. Genzel et al. \shortcite{gr98}). However in
many cases the identity of the central source is still uncertain.

A large number of ULIRGs have now been identified. Many have been well
studied via optical imaging \cite{lk94,cd96,am96,mt96,sj00}
and spectroscopy \cite{ld99,kd98,vs99a}. However, data of longer
wavelengths, such as the near infrared (NIR), have the potential to
probe deeper into the dust which enshrouds these galaxies, since the
extinction is much less i.e. A$_{V} \sim$ 10 A$_{K}$. Relative to the
optical, less NIR data exists for ULIRGs. Goldader et
al. \shortcite{gj95} have obtained K band spectra for a sample of 13
ULIRGs and 24 luminous infrared galaxies (LIRGs). Comparison of the
properties of the two sets of galaxies highlights a suppression in the
dereddened luminosity of the Br$\gamma$ emission, relative to the
far-infrared 
luminosity for the ULIRGs, as compared to the LIRGs. Murphy et
al. \shortcite{mt96} have studied a sample consisting of 46 galaxies
from the \iras \ 2 Jy survey. They have imaged these galaxies in the
optical and the NIR, and find a high percentage to be interacting
(95 per cent). They have also obtained NIR spectral data for 33 of these
objects for which results has recently been published
\cite{mt99,mt00}. Finally, Veilleux et al. \shortcite{vs97,vs99b} have 
conducted a campaign aimed at uncovering hidden broad line regions in
ULIRGs. They have observed in total 64 ULIRGs, taken from the \iras \
1 Jy sample \cite{kd95,kd98,kd98ii}. Broad Pa$\alpha$ is detected for
the first time in four sources and possibly more, some uncertainty is
due to the potential contamination from the presence of a nearby
H$_{2}$ line.  

We have selected a subset of galaxies mainly from the sample of Leech
et al. \shortcite{lk94}. These display a cross-section of interaction
characteristics i.e. ranging from some that are obviously interacting
to those with no companion within a significant distance. Two other
ULIRGs were added which show evidence of interaction in their
morphology. Our final sample shows a bias towards interacting
galaxies. 

H and K band spectra have been obtained for this sample. Using this
data we have investigated the central sources of the ULIRGs, looking
for possible broadened emission lines and/or high excitation lines
indicative of the presence of an AGN. We also study the excitation
mechanisms of the lines present. This will allow us to determine which
processes are dominant in particular objects, for example excitation
due to young stars, supernova remnants (SNR), a central black hole or
shocks due to interaction. Additionally we investigate further
relationships which have been established previously, such as the
under-luminous Br$\gamma$ line issue mentioned previously. In addition
we have J, H and K band imaging for our sample which is discussed
separately in Davies et al. \shortcite{dr00b}. 

This paper is organised as follows. First, the sample is discussed in
\S~\ref{sample}. The spectral data are then presented and the
observations and data reduction are discussed in
\S~\ref{observations}. The final results are presented and discussed
in \S~\ref{results}. The summary of our results appears in
\S~\ref{summary}, and a brief synopsis for each galaxy is included in
Appendix~\ref{app}.

\section{The sample}
\label{sample}

The galaxies in our sample were selected based on their extreme
infrared luminosities ($L_{IR} \geq 10^{12} L_{\odot}$, using H$_{o}$
= 50 km s$^{-1}$ Mpc$^{-1}$ and q$_{o}$ = 0). We note that when using
the Sanders and Mirabel \shortcite{sm96} definition of this
classification, which uses the higher value of H$_{o}$, two of our
sample, IRAS\,20210 and IRAS\,23420, have values for $L_{IR}$ falling
slightly below $10^{12}$ L$_{\odot}$. The redshift range of the
sample, z = 0.06 -- 0.21, was chosen to be well suited to ground-based
NIR observations. For all the objects Pa$\alpha$ is shifted into the K
band, except IRAS\,23498 which was only observed in the H band in
which Pa$\beta$ falls. The [FeII] line at 1.644 $\mu$m and a number of
H$_{2}$ lines also fall into the K band window.

Five of the seven objects in our sample are a small subset of a large
sample observed in the optical by Leech et al. \shortcite{lk94}. Their
original sample was complied from four flux limited (at $S_{60}$ = 0.6)
sub-samples of the QDOT redshift survey of {\it IRAS} galaxies
\cite{la99}. These consist of (1) all known ULIRGs from the North
Galactic Wedge (NGW) survey of Lawerence et al. \shortcite{la86}, (2)
all known ULIRGs from the QDOT survey in a specified RA and DEC range,
(3) other randomly selected ULIRGs from the QDOT survey and (4) a number
of high luminosity ($5 \times 10^{11} < L_{60\mu m}/L_{\odot} <
10^{12}$) galaxies from the same survey.

IRAS\,20210 and IRAS\,23365 were not part of the Leech sample but were
taken from the literature as additional, bright examples of nearby
interacting ULIRGs falling within the observing constraints. We note
that this sample is not complete in any sense.

Table~\ref{props} lists the basic properties of the galaxies presented
in this paper.  Galaxy names are given in column (1), positions in
columns (2) and (3), interaction class in column (4), redshift, z,
taken from the NASA/IPAC Extragalactic Database in column (5),
infrared luminosity in column (6) and galaxy classification taken from
the literature, if known, in column (7). References for
classifications are given in Appendix~\ref{app}.
 
\begin{table*}
\begin{minipage}{130mm}
\centering
\caption{Details for ULIRGs observed. \label{props}} 
\begin{tabular}{lcccccc}
{IRAS Name} &  
{RA (J2000)} & 
{DEC (J2000)} & 
{Interacting?}  &  
{z} & 
{$\log \frac{L_{IR}}{L_{\odot}}$} & 
{Comments} \\
%.........................
{(1)} &
{(2)} &
{(3)} &
{(4)} &
{(5)} &
{(6)} &
{(7)} \\
\hline
00150+4937 & 00h17m44.8s & +49d54m15s & Yes      & 0.148 & 12.64 & \\
16487+5447 & 16h49m47.2s & +54d42m32s & Yes      & 0.104 & 12.48 & LINER$^{1,2}$\\
17179+5444 & 17h18m54.2s & +54d41m47s & No       & 0.147 & 12.57 & Sy2$^{2}$ \\
20210+1121 & 20h23m25.9s & +11d31m31s & Yes      & 0.056 & 12.19 & Sy2$^{3,4}$ \\
23365+3604 & 23h39m01.3s & +36d21m10s & Disturbed& 0.065 & 12.45 & LINER$^{5}$\\
23420+2227 & 23h44m32.3s & +22d44m29s & No       & 0.087 & 12.09 & \\
23498+2423 & 23h52m25.0s & +24d40m09s & Yes      & 0.212 & 12.82 & Sy2$^{2}$\\
\hline
\end{tabular}

References: (1) Kim et al. \shortcite{kd98} (2) Veilleux et
al. \shortcite{vs99a} (3) Vader et al. \shortcite{vj93} (4) Perez et
al. \shortcite{pe90} (5) Veilleux et al. \shortcite{vs95}. 
\end{minipage}
\end{table*}

\section{Observations and Data reduction} 
\label{observations}

The spectral observations were performed using the United Kingdom
Infrared Telescope (UKIRT) on the nights of 7th and 8th August
1999. The long slit grating spectrometer, CGS4, was used with H and K
band filters. The orientation of the slit during observations was
chosen to be along the major axis of the galaxy or across both nuclei
if a second nucleus was known to be present. Integration times were
usually of the order of $\sim$ 20 -- 40 minutes, per object, in both
bands. Sky frames were obtained by nodding along the slit, thus
allowing continuous observation of the galaxies. To properly sample
the CGS4 spectrum in the configuration usually used for these
measurements, four individual spectra were obtained, with the spectrum
shifted by half a pixel over 2 pixels. This has the additional benefit
of allowing the removal of the effects of bad pixels in a single
row. These frames were combined to produce an overall spectrum.
Spectra were generally extracted using 5 pixels along the slit;
outside of these few pixels relatively little ($\sim$1 per cent) or no
signal was detected. The resulting spectra were fully sampled at a
resolving power of 800 (375 km s$^{-1}$) at 2 $\mu$m, with a slit
width and pixel size of 0.61 arcsec projected on the sky.

Basic data reduction was performed using IRAF and consisted of the
following steps; \begin{enumerate}
\item{Subtraction of bias frames and division by flat frames removing
the bias level present in the array and pixel-to-pixel variations.}
\item{Subtraction of sky frames from object frames. This removed sky
lines and created, effectively, two spectra on one image, one being
positive, the other negative.}
\item{Transformation of the images using a matrix produced from arc
images to remove distortion effects often found in long slit images.}
\item{Addition of positive and negative spectra, producing a final
spectrum.}
\item{Division by the spectrum of a type F (or earlier) star, observed 
at a similar air mass, correcting the spectra for instrumental
and atmospheric responses. Multiplication by a black body at the
effective temperature of the standard provided the intrinsic spectral
shape of the galaxy. Observations of flux standards reduced in the
same fashion were used to obtain the final flux calibration of 
the galaxy.}  
\end{enumerate}

Early type stars were chosen since they do not show metal features
seen in later type stars which would have masked features in the
galaxy spectra or produce false emission features after division. The
stars used do exhibit some atomic hydrogen features (e.g. Br$\gamma$
in the K band) which were removed using Lorentzian fits. 

\section{Results and Discussion}
\label{results}

\begin{figure*}
\centerline{\psfig{figure=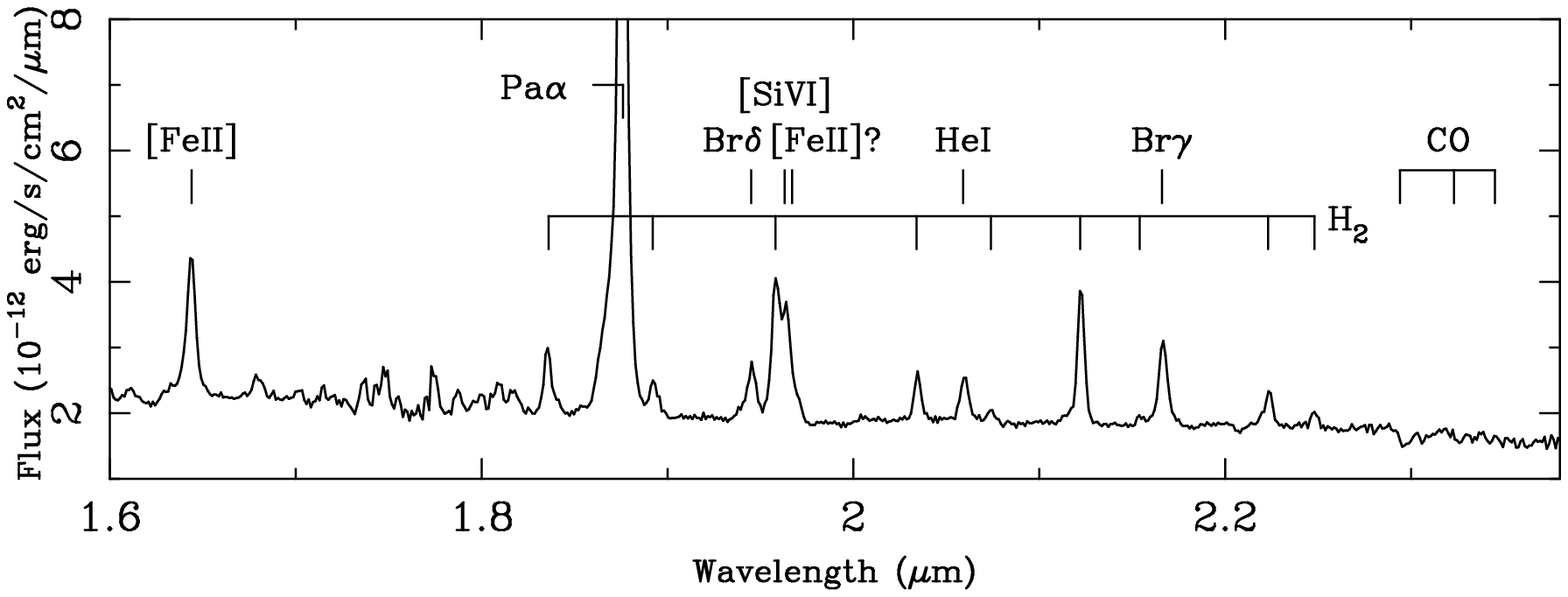,width=15.5cm}}
\caption{Merged H and K band spectrum of IRAS\,20210 with the main
lines observed labelled. The scaling has been chosen so as to enhance
the less luminous lines seen. Note that atmospheric transmission is
particularly poor for the wavelength range 1.7 -- 1.85 $\mu$m and most 
of the 'features' in this region are noise.
\label{mark}}
\end{figure*}

Table~\ref{obs} presents details of the observations made for each
object, listing the specific bands, slit position angles and on source
exposure times used. The final reduced H and K band spectra at
observed wavelengths are shown merged in
Figure~\ref{spectra}. Figure~\ref{mark} shows the merged spectrum,
produced using DIPSO, for IRAS\,20210 at rest wavelength with the
prominent lines labelled. The spectra have been scaled to the
strongest line present, Pa$\alpha$. The only exception to this is
IRAS\,23498 which was not observed in the K band. A plot of the
atmospheric transmission has been included with the plots to indicate
areas where noise may dominate the spectrum, effectively obscuring
lines present in the affected waveband. These data, produced using the
program IRTRANS4, were obtained from the UKIRT worldwide web
pages. Emission from a second nucleus was detected for IRAS\,00150 and
IRAS\,20210, these spectra have been included together with that of
the primary nucleus. In both cases the second nucleus has much lower
flux level. Note that, due to the slit position angles used, the
secondary nucleus seen for IRAS\,16487 in the K' band \cite{cd96}, and
IRAS\,23498 in the optical \cite{lk94} did not fall within the slit.

\begin{table}
\centering
\caption{Bands and slit position angles used for observations. \label{obs}}
\begin{tabular}{lllccc}
{Galaxy} &  \multicolumn{2}{c}{Band} & {PA }&  \multicolumn{2}{c}{Exp. time} \\
 & & & & \multicolumn{2}{c}{(minutes)}\\
\cline{2-3} \cline{5-6}\\
(IRAS)  & H & K & (east of north) & H & K \\
\hline
 00150 & yes          & yes          & 172 & 24 & 40 \\
 16487 & yes$^{\dag}$ & yes          & 135 & 24 & 40 \\
 17179 & yes          & yes          & 57  & 40 & 56 \\
 20210 & yes          & yes          & 166 & 24 & 40 \\
 23365 & yes          & yes          & 0   & 40 & 40 \\
 23420 & yes          & yes$^{\dag}$ & 42  & 40 & 48 \\
 23498 & yes          & no           & 108 & 24 & -- \\
\hline

\end{tabular}

$\dag$ Variable weather conditions during this observation makes flux
calibration uncertain, see main text for more details. 
\end{table}

\begin{figure*}
\centerline{\psfig{figure=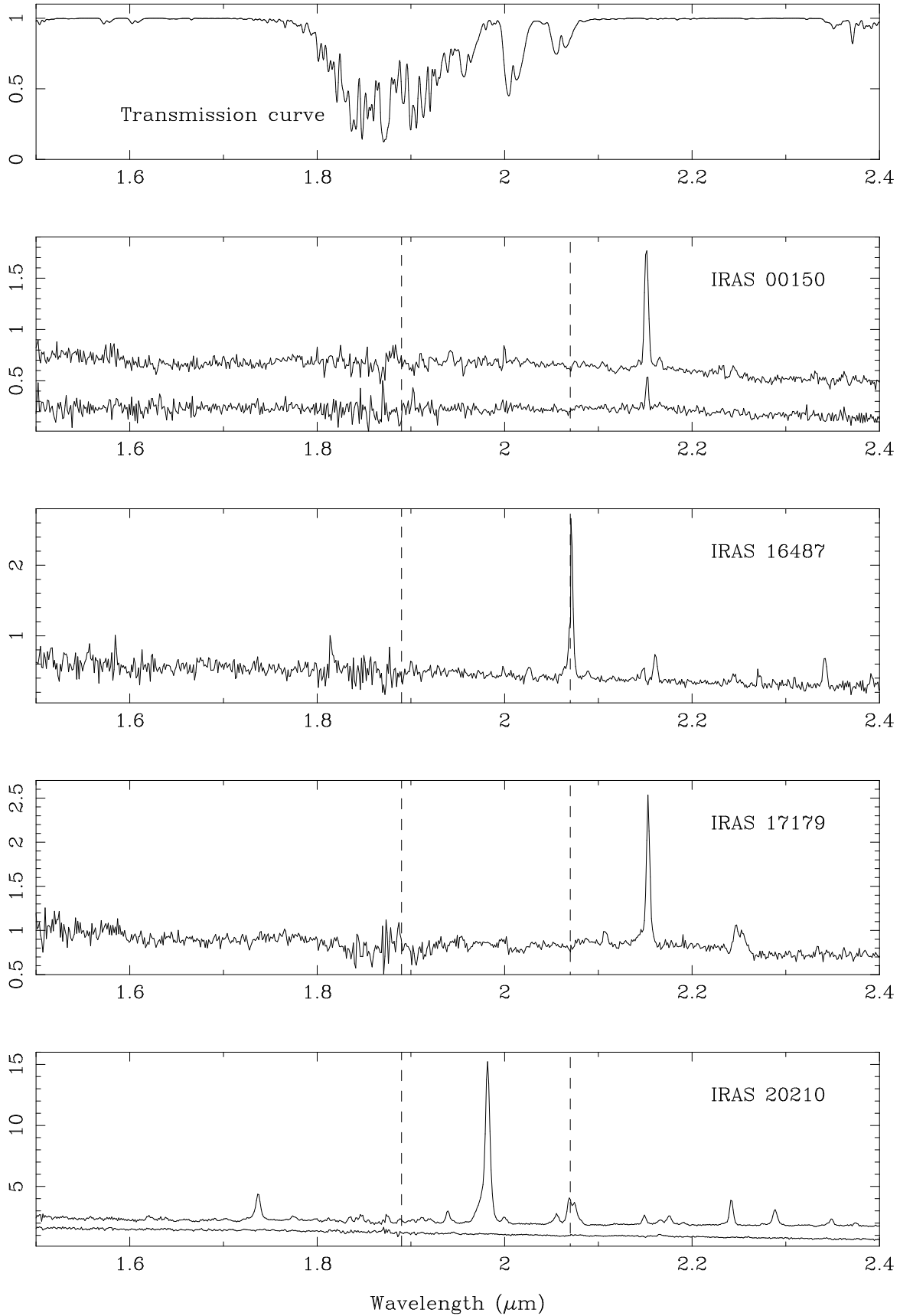,width=15.4cm}}
\caption{Merged H and K band spectra of flux at observed wavelength in 
units of $10^{-12} erg \ s^{-1} cm^{-2} \mu m^{-1}$. Each spectrum has 
been scaled to the strongest line present, Pa$\alpha$. This is not the 
case for IRAS\,2349 observed only in the H band. The vertical dashed
lines show the overlap between the H and K band observations. The
uppermost plot shows the atmospheric transmission for the same
waveband in order to indicate wavelengths where atmospheric effects
are important. These data, produced using the program IRTRANS4, were
obtained from the UKIRT worldwide web pages. For line identifications
see Figure~\ref{mark}. 
\label{spectra}}
\end{figure*}

\begin{figure*}
\centerline{\psfig{figure=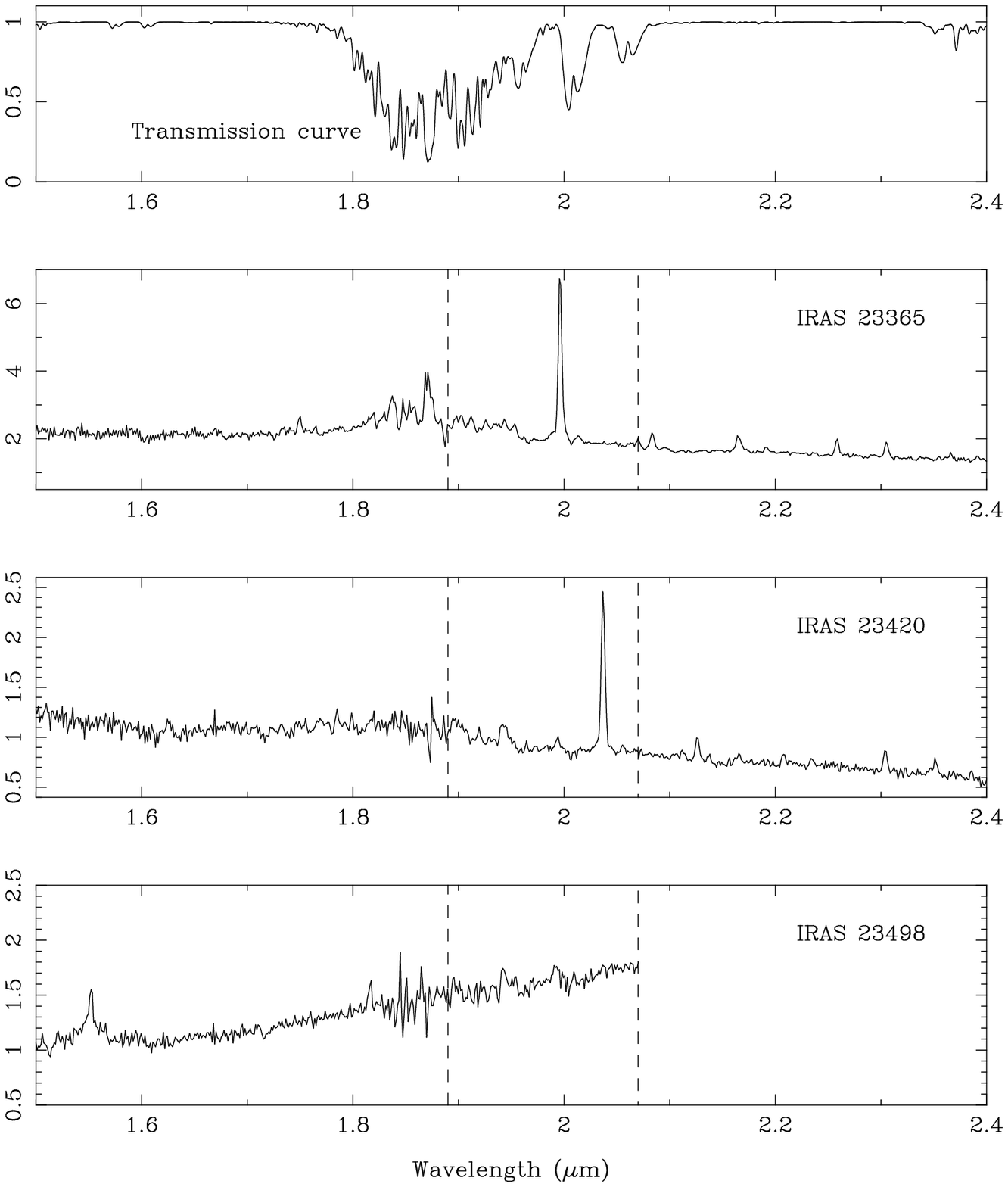,width=15.4cm}}
\contcaption{}
\end{figure*}

A useful check on the validity of the flux calibration was possible
since the H and K bands were reduced independently for each
object. The agreement between the bands is generally good, and lies
within 20 per cent. Variability in weather conditions made flux calibration
uncertain on the two occasions indicated in Table~\ref{obs}. In these
cases the band for which the calibration was uncertain was scaled to
match the other band, for the same galaxy, for which we are confident
of the flux  calibration.

Table~\ref{lines} lists the line fluxes and equivalent widths
measured using the separate H and K band spectra. In order to
accurately measure line fluxes a continuum was fitted to each spectrum and
subtracted. Points free of known spectral features were used. However,
in cases where significant variability in atmospheric conditions made
removal of atmospheric effects uncertain, such as for IRAS\,23365 in
the H band, fitting of a continuum was somewhat subjective. In order
to assess the magnitude of this effect, several low order fits were
tried, subsequent line fluxes varied by less than $\sim$10 per cent. Due to
the overlap in wavelength between the H and K bands some lines appear
in both spectra e.g. Pa$\alpha$ is visible in both bands for
IRAS\,20210, IRAS\, 23365 and IRAS\,23420. When this is the case the
final values for flux and equivalent width quoted are averages, except
in the cases mentioned above where flux calibration is uncertain due
to variable weather conditions, when the more reliable measurement was
adopted. 

\begin{table*}
\begin{minipage}{160mm}
\centering
\caption{Fluxes and equivalent widths for emission lines seen in
spectra. \label{lines}} 
\scriptsize
\begin{tabular}{lcccccccccccccc}

% ..........................................
Galaxy                  & % 1
[FeII]                  & % 2
Br9                     & % 3
H$_{2}$                 & % 4
Pa$\alpha$              & % 5
H$_{2}$                 & % 6
Br$\delta$              & % 7
H$_{2}$                 & % 8
[SiVI]                  & % 9
H$_{2}$                 & % 10
HeI                     & % 11
H$_{2}$                 & % 12
Br$\gamma$              & % 13
H$_{2}$                 & % 14
H$_{2}$                \\ % 15
% ..........................................
(IRAS)                  & % 1 
1.644                   & % 2
1.818                   & % 3 
1.836                   & % 4 
1.876                   & % 5 
1.892                   & % 6 
1.945                   & % 7 
1.958                   & % 8 
1.963                   & % 9
2.034                   & % 10
2.059                   & % 11
2.122                   & % 12
2.166                   & % 13
2.223                   & % 14
2.248                   \\% 15
% ..........................................
(1) & (2) & (3) & (4) & (5) & (6) & (7) & (8) & (9) & (10) & (11) &
(12) & (13) & (14) & (15) \\ 
% ..........................................
\hline
00150a           &  ...      &   ...    & ...      &
     0.671      & ...       & 0.016    & ...      & ...      &
     0.011$^{x}$& ...     & ...      & 0.017    &
                &           \\ 
                &  ...      &  ...     & ...      &
     92.6       & ...       & 8.4      &  ...     & ...      &
     6.7       & ...      & ...      & 13.8     &
                &           \\ 
00150b           &  ...      & ...      & ...      &
     0.108      & ...       & ...      & ...      & ...      &
     ...      & ...       & ...      & ...      &
                &           \\ 
                &  ...      & ...      & ...      &
     50.1       & ...       &  ...     & ...      & ...      &
     ...      &  ...      & ...      & ...      &
                &           \\ 

16487$^{\dag}$&\it{0.256}& ...     & 0.074    &
     0.938      &0.037      &    0.067 & 0.160   & ...      &
     0.058      &0.040      & 0.182    & 0.06    &
                &           \\ 
                &\it{0.4 }  &  ...     & 17.0     &
     216.6      & 9.1       & 17.7     & 42.6     & ...      &
     17.3       & 12.6      & 62.0     & 20.7     &
                &           \\ 

17179 &\it{0.050}& ...       &0.060     &
     0.773      & ...       & ...      & 0.146    & 0.110    &
     0.022$^{x}$&\it{0.018$^{x}$}& 0.092    &0.056
                &           \\ 
      &\it{19.8}& ...       & 6.9      &
     90.9       & ...       & ...      & 18.9     & 14.3     &
     2.9$^{x}$  &2.4$^{x}$  & 13.0     &7.9       &
                &           \\ 
20210a     & 1.237    &\it{0.144}&\it{0.382}&
     9.371      & 0.317     & 0.631    & 1.241    & 0.860    &
     0.317     & 0.404    & 0.943    & 0.686    &
     0.236      &0.121      \\ 
                & 54.7     &\it{7.1}  &\it{18.3} &
     46.2       & 15.9     & 32.4     & 63.4     & 42.8     &
     16.6     & 21.3     & 50.2     & 36.9     &
     12.9       &6.8        \\ 

20210b             &  ...     & ...      & ...      &
    ...         & ...       &  ...     & ...      & ...     &
    ...         & ...       & ...      & ...      &
    ...         & ...       \\ 
                &  ...     & ...      & ...      &
    ...         & ...       &  ...     & ...      & ...      &
    ...         & ...       & ...      & ...      &
    ...         & ...       \\ 
23365            & 0.163     & ...      & ...      &
     2.266      & 0.096     & 0.068    & 0.186    & ...      &
     0.073      & 0.068     & 0.173    & 0.157    &
     0.051      & ...       \\ 
                & 7.4       & ...      & ...      &
     115.8      & 6.9       & 3.8      & 10.8     & ...      &
     4.5        & 4.3       & 11.3     & 10.5     &
     3.6        & ...       \\ 
23420$^{\ddag}$  & 0.126     & ...      & 0.069    &
     0.844      & 0.057     & 0.036    & 0.115    & ...      &
     0.045      & 0.035     & 0.111    & 0.079    &
     ...        & ...       \\ 
                & 12.6      & 3.3      & 7.5      &
     81.8       & 4.9       & 3.3      & 10.5     & ...      &
     4.4       & 3.5      & 12.1     & 9.2      &
     ...        & ...       \\ 
23498      &0.100$^{x}$&     &          &
                &           &          &          &          &
                          &           &          &          &
                &           \\ 
                &6.1$^{x}$  &          &          &
                &           &          &          &          &
                          &           &          &          &
                &           \\ 
\hline

\end{tabular}
\footnotesize

\dag Variable weather conditions during the exposure of IRAS\,16487 in
the H band make flux calibration uncertain. 

\ddag Variable weather conditions during the exposure of IRAS\,23420
in the K band make flux calibration uncertain. 

(x) Indicates line detected only at 2$\sigma$ limit, not above
3$\sigma$ as is the standard. 

Columns: 
(1) Name of object. 
(2) -- (15) list the fluxes and equivalent widths for the major lines
observed where the top value is observed line flux, F (10$^{-14}$ erg
s$^{-1}$ cm$^{-2}$) and bottom value is observed equivalent width, EW
(\AA). A value in italics indicates the lines falls in a waveband
where atmospheric transmission falls below 80\%, ... indicates a
non-detection and a blank column indicates the line is at a wavelength
redshifted out of the observed band.  
\end{minipage}
\end{table*}

In the majority of the spectra a blue 'shoulder' is evident on the
Pa$\alpha$ line ($\lambda1.876\mu$m). The origin of this will be
discussed further in \S~\ref{h1lines}. All fluxes and equivalent
widths quoted in Table~\ref{lines} for Pa$\alpha$ include this
'shoulder'. Of note is a line blend consisting of contributions from
Br$\delta$ ($\lambda1.945\mu$m), H$_{2}$ 1--0S(3) ($\lambda1.958\mu$m)
and [SiVI] ($\lambda1.963\mu$m), prominent in the K band for all
objects. To facilitate flux measurements these lines were deblended
using the routine 'splot' from IRAF and also using DIPSO.

Note that Pa$\beta$ was observed in the H band for IRAS\,23498. 
Although not given in Table~\ref{lines} this line is discussed
further in \S~\ref{h1lines}.  

\subsection{General features of the spectra}

As can be seen the K band spectra are much richer in lines than the H
band for the redshifts of the galaxies observed. In the K band most of
the galaxies observed show several hydrogen recombination lines, most
notably Pa$\alpha$ and Br$\gamma$. A number of H$_{2}$ excitation
lines are seen, especially numerous for the brighter galaxies
i.e. IRAS\,20210. Other lines visible are HeI at 2.059$\mu$m and [FeII]
at 1.644$\mu$m. [SiVI] at 1.963$\mu$m is observed for IRAS\,17179 and
IRAS\,20210.

Several CO bandheads longward of 2.3$\mu$m can be seen in two of the
galaxies observed at this wavelength: IRAS\,20210 and IRAS\,23365. These
are not obvious in Figure~\ref{spectra} due to the scaling used. See
\S~\ref{co} for further discussion. 

The H band shows few features for the majority of galaxies observed,
with the exception of IRAS\,23498 which shows a clearly broadened
Pa$\beta$ line with FWHM $\sim$ 5000 km s$^{-1}$. This is discussed
further in \S~\ref{h1lines}. 

\subsection{Extinction}
\label{ext}

Values of internal reddening were calculated using the NIR
recombination lines, measured here, and optical lines taken from the
literature, available for four of the seven galaxies in this sample.
Table~\ref{extinc} lists the NIR - optical line combinations used and
extinction values found for each. The method used follows that of
Veilleux and Osterbrock \shortcite{vo87}, using the reddening curve
for the optical and NIR parametrised by Cardelli et
al. \shortcite{cj89}. The intrinsic H$\alpha$/H$\beta$ ratio was taken 
to be 2.85 for starburst galaxies (assuming Case B recombination) and
3.1 for AGN (having enhanced H$\alpha$ due to collisional excitation
\cite{vo87}). Galactic extinction for each galaxy is shown in
Table~\ref{extinc}. These values are small and can be assumed to be
negligible, even for IRAS\,20210, which is closest to the galactic
plane with a galactic latitude of -14$^{o}$ and A$_{V}$ of 0.50
mag. We would expect values for A$_{V}$ using only optical lines to
be less than those derived from IR lines, since optical lines probe to
a  smaller optical depth than those in the NIR \cite{pp91}.

\begin{table*}
\begin{minipage}{100mm}
\centering
\caption{Values found for extinction using various line ratios in
the optical and infrared. Note that IRAS\,00150 and IRAS\,23420 
have no optical observations available in the literature and hence
have not been included here.  
\label{extinc}}
\begin{tabular}{lccccccc}
{} & \multicolumn{6}{c}{A$_{V}$ (mags)}  & {}\\
\cline{2-7} \\
{Galaxy} & 
{$\frac{H\alpha}{H\beta}$} & 
{$\frac{Pa\alpha}{H\beta}$} & 
{$\frac{Br\gamma}{H\beta}$} & 
{$\frac{Pa\alpha}{H\alpha}$} & 
{$\frac{Br\gamma}{H\alpha}$} &
{Galactic$^{\dag}(A_{K})$} &
{Ref$^{\ddag}$}\\
(IRAS) & & & & & & & \\
\hline
 16487 & 1.49 & 3.01 & 1.33 & 3.78 & 1.24 & 0.06(0.01) & (1)\\
 17179 & 3.72 & 4.81 & 3.81 & 5.36 & 3.88 & 0.09(0.01) & (1)\\
 20210 & 2.54 & 3.19 & 2.26 & 3.53 & 2.08 & 0.50(0.06) & (2)\\
 23365 & 2.79 & 4.99 & 3.66 & 6.14 & 4.09 & 0.34(0.04) & (3)\\
\hline
\end{tabular}

\dag Taken from Schlegel et al.\shortcite{sd98}

\ddag Source from literature used for optical line ratio.

References: (1) Veilleux et al. \shortcite{vs99a}, (2) Young et
al. \shortcite{ys96}, (3) Veilleux et al. \shortcite{vs99b}.

\end{minipage}
\end{table*}

Potential problems with the use of this method exist. It is strictly
only valid if the dust in these galaxies, known to be an important
component in ULIRGs, acts as a discrete absorbing screen in front of
the emitting gas. If mixed with the gas, dust scattering may have an
important effect, preferentially enhancing H$\beta$
emission. Absorption features from the underlying stellar continuum
may have affected emission lines for some objects. Observing effects
such as wavelength dependent slit losses, seeing, guiding effects and
centring errors may affect the NIR line fluxes, especially important
here since most of the objects are extended. Values for H$\alpha$ and
H$\beta$ fluxes come from observations using a wider slit than used
here i.e. Veilleux et al. \shortcite{vs99a} used a 2.0 arcsec slit,
our observations were made with a 1.22 arcsec slit width. This could
lead to underestimates of NIR line fluxes and hence A$_{V}$, in cases
where the emission line region is extended over many arcsecs. Finally,
differences in the dust coverage could affect these values, potential
'clumpy' dust coverage may result in NIR lines coming from a region
not seen in the optical. Even given all these uncertainties, it is
expected that general trends in line ratios will still be revealed.

\subsection{HI recombination lines}
\label{h1lines}
\subsubsection{Pa$\alpha$ emission}
Several Paschen and Brackett series recombination lines are evident in
these spectra. The brightest line for each galaxy is Pa$\alpha$, at a
rest wavelength of 1.876$\mu$m. In every observation Pa$\alpha$ is
narrow, having deconvolved FWHM $\sim$ 300 -- 500 km\,s$^{-1}$, except
for IRAS\,20210 which shows FWHM $\sim$ 700 km\,s$^{-1}$. Notably, in
all but IRAS\,00150, a blue asymmetric wing at Pa$\alpha$ is seen to
some degree. This phenomenon has been noted in ULIRGs previously
e.g. Veilleux et al. \shortcite{vs97} who find that many of their
objects display blue asymmetric wings at Pa$\alpha$, including
IRAS\,17179 and IRAS\,23498. This could be faint broad emission from
high velocity gas (FWHM $\sim$ 2500 km s$^{-1}$). However in most
cases, with the possible exception of IRAS\,17179 and IRAS\,20210,
other explanations are possible. One possibility is faint H$_{2}$
emission from the 7--5O(3)1.8721$\mu$m and 6--4O(5) 1.8665$\mu$m
transitions \cite{vs97,vs99b}. Another possibility is the presence of
a pair of HeI emission lines at 1.8686$\mu$m and 1.8697$\mu$m
\cite{mt99}. At the spectral resolution of our data we cannot
distinguish between the possibilities.

\subsubsection{Broadened emission}
The only broad emission observed is Pa$\beta$ with FWHM $\sim$~5000 km
s$^{-1}$, seen in the spectrum of IRAS\,23498. Previous observations
of this Seyfert 2 consistently show broadened emission (FWHM
$\sim$~3000 km s$^{-1}$) in the Pa$\alpha$ line \cite{vs97}.

\subsubsection{L$_{Br\gamma}$ vs L$_{FIR}$ correlation} 
\label{corr}

A significant correlation between L$_{Br\gamma}$ and L$_{FIR}$ has
been found for LIRGs, the less luminous counterparts of ULIRGs. ULIRGs
however, do not follow the same relationship and appear to be
underluminous in Br$\gamma$ emission \cite{gj95}. We have added the
data for our sample to the previously published data to see if they
fall upon the same relationship.

For consistency we have calculated L$_{FIR}$ ($\equiv$
L(40--120$\mu$m)) using the formula from Helou et
al. \shortcite{hg85}. Note that, following Goldader et
al. \shortcite{gj95}, we have taken H$_{o}$ = 75 km s$^{-1}$
Mpc$^{-1}$ for this calculation. Where available, we have used {\it
E(B--V)} from the literature, calculated using optical line ratios
(see Table~\ref{extinc} for more details). Where {\it E(B--V)} was not
available the data were not corrected for extinction.

\begin{figure}
\centerline{\psfig{figure=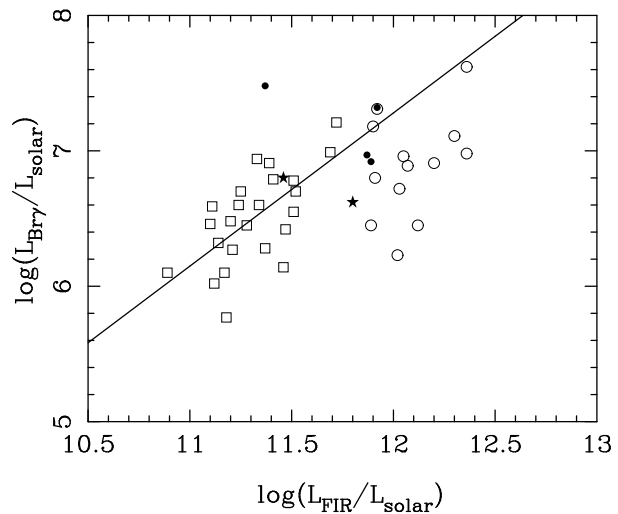,width=8cm}} 
\caption{Dereddened Br$\gamma$ luminosity vs far infrared
luminosity. Filled circles represent ULIRGs from this sample, filled
stars represent ULIRGs from this sample uncorrected for
extinction. Empty squares and circles represent LIRGs and ULIRGs
respectively from the sample of Goldader et al. \shortcite{gj95}. The
solid line is the best fit to the LIRG sample, also from Goldader et
al. \shortcite{gj95}.
\label{gold}} 
\end{figure}

Figure~\ref{gold} shows L$_{Br\gamma}$ vs L$_{FIR}$ for both datasets, 
where the solid line is the best fit to the LIRGs made by Goldader et
al.. They conclude that the ULIRGs from their sample fall below the
correlation indicating that they are underluminous in Br$\gamma$. The 
two lower luminosity galaxies from our sample, classified in this work
as ULIRGs, would actually be classified in the Goldader scheme as
LIRGs, due to the different value of H$_{o}$ used. Our higher
luminosity objects do not seem to follow the trend of underluminosity
in Br$\gamma$ emission with respect to the correlation marked. In fact,
our data points may follow the correlation if the large scatter seen
for the LIRGs is taken into consideration.

The importance of aperture effects must also be considered. On
average, the objects from the Goldader sample have been observed with
a 3.0 arcsec wide aperture. The slit width used in our observations
was 1.22 arcsec. This may result in our observations giving a lower
line flux, and hence luminosity, by a factor of approximately 2, with 
respect to those of Goldader et al..

The depression of the Br$\gamma$ luminosity in ULIRGs, when compared
with LIRGs, is explained by Goldader and colleagues by a simple model
where ULIRGs contain very compact nuclei which are optically thick at
2$\mu$m. They reason that the Br$\gamma$ is therefore highly obscured,
resulting in a poor correlation between Br$\gamma$ and FIR luminosity. 

An alternative explanation would be to question the correlation found
for LIRGs. When considered as a group, the relationship found can be  
seen to be very dependent on the two highest luminosity points. If
these are disregarded, then the scatter of the remaining points 
would change the fitting of any trend dramatically. When the
additional higher luminosity data points provided by the ULIRGs are 
added to the plot then a correlation with a lower gradient could be
fitted to both sets of data. Indeed, there is such a large scatter in
the data that it is possible that there is no correlation and instead
simply two 'regions', representing the range of properties found for
each group. They find support for their conclusions in other work
involving LIRGs, where similar relationships have been found
\cite{dd87,vs95}. The distinction between LIRGs and ULIRGs is
arbitrary, and claims that they are fundamentally different would need 
to be supported by observations at other wavelengths.

\subsubsection{Second nuclei}

Four of the galaxies observed have multiple nuclei. Of these, two were
observed with the slit orientation covering both nuclei. This was the
case for IRAS\,00150 and IRAS\,20210. The spectra of the second nuclei are
plotted with the primary nuclei, at the same scaling, in
Figure~\ref{spectra}. As can be seen, the spectra of the secondary 
nuclei seen are practically featureless. This is consistent with
preliminary findings from another recent NIR survey of 33 ULIRGs
(including IRAS\,16487 and IRAS\,23365) conducted by Murphy et
al. \shortcite{mt99}. 

The only feature we see for the secondary nucleus of IRAS\,00150 is
weak Pa$\alpha$ emission. The second nucleus of IRAS\,20210 shows no
emission lines, but does show CO absorption, indicative of an old
stellar population. It is useful to make an estimate of the
contribution of the secondary nucleus to the overall FIR luminosity of
the system as detected by \iras. For IRAS\,00150, assuming the
relationship between L$_{Br\gamma}$ and L$_{FIR}$ holds
(\S~\ref{corr}) and using the ratio of Pa$\alpha$ in the two nuclei,
we estimate the minor nucleus contributes 20 per cent of the FIR
emission. For IRAS\,20210 we used the ratio of Br$\gamma$ emission.
An upper limit of F$_{Br\gamma}$ = 3.81 $\times 10^{-16} erg \ s^{-1}
cm^{-2}$ was found in the secondary nucleus, indicating a contribution
of $\sim$ 6 per cent. Therefore, in both cases, the contribution of
the secondary nucleus is minor. We note the the similarity of the
spectra of the two nuclei seen for IRAS\,00150, here the minor
component could be a scaled version of the major. This is not the case
for IRAS\,20210, where the line to continuum ratio seen in the major
component is not mirrored in the second nucleus.

\subsection{Excitation of H$_{2}$ lines}

\subsubsection{Thermal vs non-thermal emission}

Emission from electric quadrupole transitions within the ground state
of molecular hydrogen is often bright in infrared observations of
starburst galaxies and AGN. There are two types of excitation
mechanisms producing these lines: collisions due to shocks (thermal
process) and radiative decay from an excited state produced by UV
photons or X-rays, resulting in fluorescence (non-thermal process). In
general both mechanisms, in varying proportions, contribute to the
emission. Two galactic cases of particular interest are (1) supernova
remnants (SNR) which exhibit purely thermal shock excited emission,
and (2) planetary nebulae which exhibit pure fluorescent emission. In
starburst galaxies OB stars produce UV photons, also SNR and winds
from massive stars drive large scale outflows which produce shocks
(e.g. Heckman et al. \shortcite{ht90}. Often the more dominant
mechanism is uncertain. In AGN, the stronger H$_{2}$ emitters, the
emission is believed to be associated with the active nucleus which
emits a strong UV -- X-ray continuum and also drives outflows.

Enhanced H$_{2}$ emission in mergers may arise from shocks produced by
cloud-cloud collisions, as originally proposed by Harwit et
al. \shortcite{hw87}. However the small filling factor of the dense
molecular clouds presents problems for this model. A solution to this
problem may be that proposed by Jog \& Solomon \shortcite{jc92}, who
describe a scenario in which HI clouds, associated with the
interacting galaxies and with large filling factors, collide.  This
leads to the formation of hot, ionised, high pressure gas which causes
a radiative shock compression of the denser H$_{2}$ clouds.

Since the non-thermal mechanism is more efficient than the thermal in
the population of $\nu$ = 2 and higher vibrational levels, in
principle line ratios can be used to discriminate between these
mechanisms.

A minimum of two H$_{2}$ lines were detected for each galaxy,
increasing to nine for the brightest galaxy observed, IRAS\,20210. The
potential for deconvolution of this emission into thermal and
non-thermally excited components is small for the galaxies which show
only a few H$_{2}$ lines and so only generalised conclusions can be
drawn. For the brighter galaxies: IRAS\,20210, IRAS\,23365 and
IRAS\,23420 we use the population and diagnostic diagrams discussed
later in this section.

\subsubsection{Line ratios: a direct comparison}

First we draw some general conclusions concerning thermal vs
non-thermal excitation for each galaxy by the direct comparison of
observed line ratios with those from models. Table~\ref{ratios}
lists the measured ratios compared with typical shock and low density
fluorescent predictions taken from Black and van Dishoeck
\shortcite{bj87}, plus a mixed model assuming equal contributions to
1--0S(1) flux from shock and UV components. 

IRAS\,00150, IRAS\,16487 and IRAS\,17179 have only two H$_{2}$ line
detections. This limits our analysis substantially, but we can
conclude that the ratio is indicative of a shock origin over
fluorescence. We note that in IRAS\,00150 and IRAS\,17179 the ratio is
$\sim$10 per cent less than that expected for shock
excitation. However this may be attributed to errors in the marginal
detection of the lines above the noise in the spectra.

Line ratios for IRAS\,20210 strongly indicate a purely shock origin
for the emission with good agreement for all but one of the ratios
considered. The one discrepancy is the 2-1S(1) line which was only
detected above the 2$\sigma$ level.

The ratios measured for IRAS\,23365 are fitted best by the predictions
produced by the mixed model. This predicts similar fluxes for 1--0S(0)
and 2--1S(1), indicating approximately equal contributions from shock
and fluorescent excitation processes.

Finally, considering the ratios seen for IRAS\,23420 the best model is
unclear. The 1--0S(2) line ratio falls between predicted values for
both the shock and mixed models. The low ratio value seen for 1--0S(0)
however, the stronger of the two lines, favours the shock model over
the mixed model.

\begin{table*}
\begin{minipage}{130mm}
\centering
\caption{H$_{2}$ line ratios compared with predicted ratios. \label{ratios}}
\begin{tabular}{ccccccccccc}
%.........................
{Line} &  
{Wavelength} & 
\multicolumn{6}{c}{Relative flux of IRAS galaxies$^{\dag}$} &
\multicolumn{3}{c}{Models$^{\ddag}$}\\
\cline{3-8}\\
%.........................
{} & 
{$\mu$m} & 
{00150} & 
{16487} & 
{17179} & 
{20210} & 
{23365} & 
{23420} &
{Shock} &
{UV} &
{Mixed} \\
%.........................
\hline
1-0S(2) & 2.0338 & 0.25 & 0.32 & 0.24 & 0.34 & 0.42 & 0.41 & 0.37 & 0.50 & 0.44 \\
2-1S(3) & 2.0735 &      &      &      & 0.07 &      &      & 0.08 & 0.35 & 0.22 \\
1-0S(1) & 2.1218 & 1.00 & 1.00 & 1.00 & 1.00 & 1.00 & 1.00 & 1.00 & 1.00 & 1.00 \\
1-0S(0) & 2.2235 &      &      &      & 0.25 & 0.29 & 0.15 & 0.21 & 0.46 & 0.34 \\
2-1S(1) & 2.2477 &      &      &      & 0.13 & 0.29 &      & 0.08 & 0.56 & 0.32 \\
\hline
\end{tabular}

\dag Normalised to the flux of 1-0S(1) emission line.

\ddag Respectively model S2 and model 14 taken from Black and van
Dishoeck \shortcite{bj87}. The mixed model assumes 50\% of 1-0S(1)
emission is thermal and 50\% is non-thermal. 

\end{minipage}
\end{table*}

As is clear from the above, it is not possible to attribute the
emission uniquely for half of the sample observed. Throughout, we are
constrained by the lines observed which mostly come from the $\nu$=1
levels. The population of these levels is dominated by thermal
excitation which falls off rapidly in other bands. Hence to deconvolve
this emission successfully, detection of lines from the $\nu$=2 levels
is necessary. This becomes evident when comparing the predicted ratios
in Table~\ref{ratios}. The best lines to use are those having very
different ratios for shock and UV models. The 1--2S(1) line, which has
been detected in every galaxy in this sample, does not fulfill this
criteria, having only a $\sim$10 per cent difference between the
models. 2--1S(3) and 2--1S(1) having $\sim$30 per cent and 50 per cent
change in predicted values are better. Unfortunately they are the
weaker lines and are often undetected.

\subsubsection{Population and diagnostic diagrams}

Various methods exist to discriminate further between thermal and
non-thermal excitation mechanisms. We use two of these methods for the 
brighter objects in our sample (IRAS\,20210, IRAS\,23365 and IRAS\,23420) 
in order to better constrain the excitation mechanisms. Level
population diagrams are a useful tool, employing similar theory to the 
previous method, but utilising more emission lines. We also employ
diagnostic diagrams with H$_{2}$ line ratios, used by Mouri
\shortcite{mh94} to study different thermal excitation mechanisms
using characteristic temperatures.

\begin{figure}
\centerline{\psfig{figure=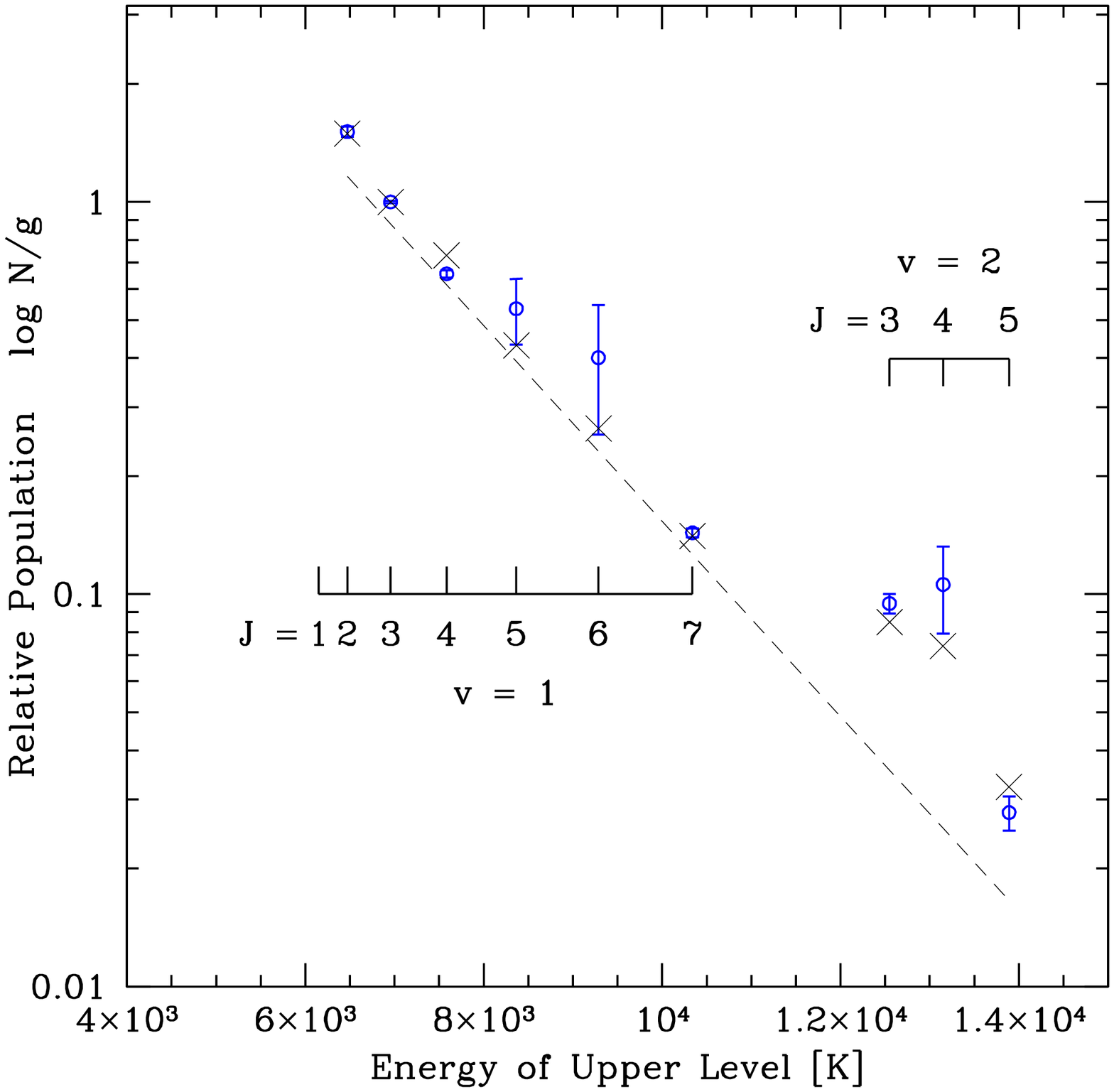,width=7.5cm}}
\centerline{\psfig{figure=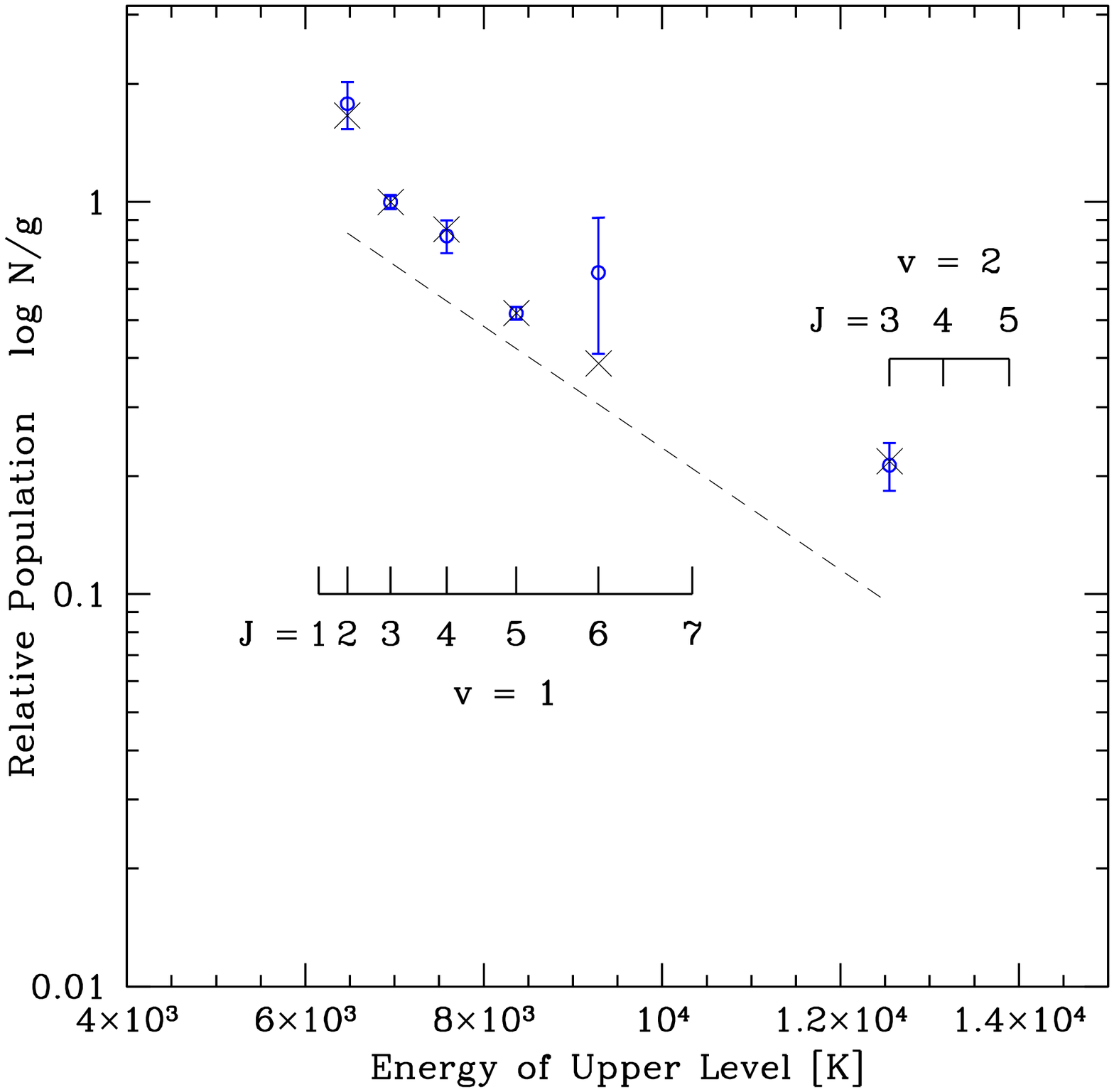,width=7.5cm}}
\centerline{\psfig{figure=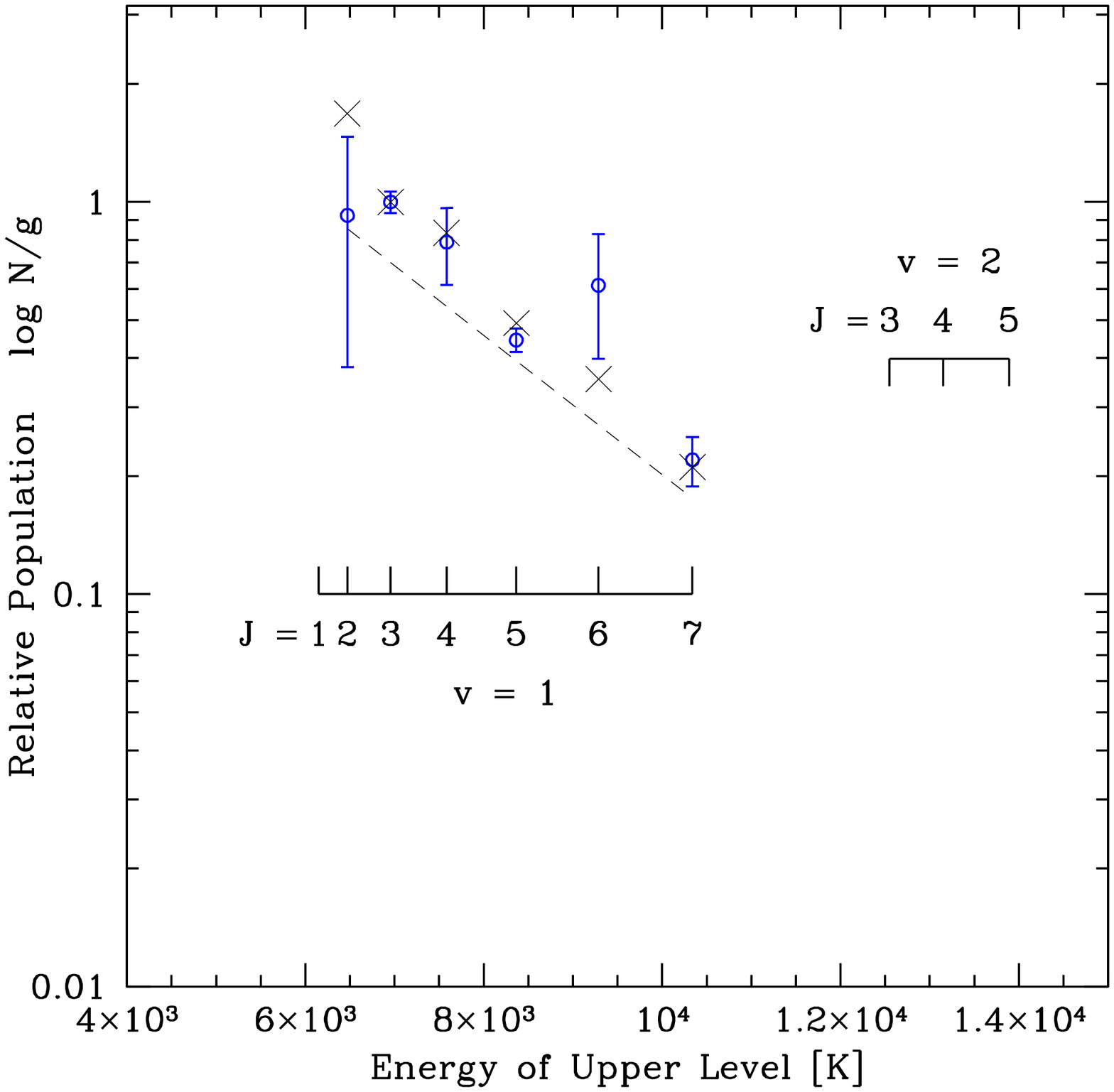,width=7.5cm}}
\caption{Relative population diagrams for IRAS\,20210, IRAS\,23365
and IRAS\,23420, from top to bottom, derived for H$_{2}$ lines
detected. The crosses denoted the combined fit to the data of a
thermal model (dotted line) and a non-thermal model (model 14, Black
and van Dishoeck \shortcite{bj87}).
\label{pop_diag}}
\end{figure}

Level population diagrams for IRAS\,20210, IRAS\,23365 and IRAS\,23420
are shown in Figure~\ref{pop_diag}, derived using the local thermal
equilibrium ortho/para ratio of 3. The level population is the column
density, N$_{u}$, divided by the statistical weight, g$_{u}$, and is
calculated using the relationship: 
\[
\frac{N_{u}}{g_{u}} \propto \frac{I}{A_{ul}\nu},
\]
using the measured flux, I, transition probability, A$_{ul}$, and
frequency, $\nu$. The statistical weight, g$_{u}$, is given by the
product of the rotational and spin degeneracies. The values of level
population are scaled to that of the $\nu$=1, J=3 level.

Simplistically, if the points on such a diagram can be fitted by a
straight line then the emission mechanism is purely thermal, otherwise
there is additional emission due to non-thermal mechanisms. However
data needs to be available for both $\nu$=1 and 2 levels, as a
straight line fit using only $\nu$=1 lines can be obtained for a
galaxy even when a significant amount of fluorescent emission is
present, as noted by Tanaka et al. \shortcite{tm89} for NGC\,7027 and
M1-78, which exhibit 10 per cent and 20 per cent non-thermal emission
respectively.  

Following the method used by Davies et al. \shortcite{dr00a}, a
combination of thermal and non-thermal models have been fitted. Free
parameters were the relative fraction of non-thermal emission,
$f_{UV}$, the temperature of the thermal emission, $T$, and the
absolute scaling. The thermal component was represented by a Boltzmann 
distribution at temperature T given by
\[
\frac{N_{u}}{g_{u}} \propto \exp \frac{-E_{u}}{kT},
\]
where $E_{u}$ is the excitation energy of the upper level. The
non-thermal model used was model 14 from Black and van Dishoeck
\shortcite{bj87}. The best fit was found using a chi-square
minimalisation technique. 

On scrutiny of Figure~\ref{pop_diag} it is immediately obvious that
none of the galaxies exhibits purely thermal emission. The data for
IRAS\,20210, with 9 derived points, indicates a {\it mostly} thermal
origin with a small fraction of fluorescence also present. Analysis
indicates 10 per cent of the 1-0S(0) flux, 40 per cent of the total H$_{2}$ flux,
comes from non-thermal emission, with a temperature of 1870K. This
model fits well for $\nu$=1 lines but fits less well for the $\nu$=2
lines. Additionally, when placed on the diagnostic diagrams of Mouri
\shortcite{mh94}, IRAS\,20210 is in the region occupied by the shock
model and supernova remnants -- objects known to show purely thermal
emission. 

The population diagram for IRAS\,23365 is quite different, a single
straight line fit is very unlikely. The lines seen for $\nu$ = 1, J =
6 and 7 and $\nu$ = 2, J = 3 are weak detections and in one case (J=7)
only an upper limit, which was not included in the fit. We find for
this object a higher proportion of non-thermal contribution, 30 per cent of
the 1--0S(1) flux, 70 per cent of the total H$_{2}$ flux, with a
temperature of 2800K. This model fits well except for the
1--0S(4) line which falls at a wavelength which shows a drop in
transmission for this object. When plotted on the diagnostic diagrams
of Mouri \shortcite{mh94}, this galaxy falls in the region occupied by
NGC\,6240. NGC\,6240 is a galaxy-merger starburst/LINER system with
much disputed emission mechanisms. Values for the amount of
non-thermal H$_{2}$ emission exhibited by this object range from none
\cite{sh97} to 70 per cent \cite{tm91}. Mouri \shortcite{mh94} favours shock
heating for the thermal component of this galaxy. Based on the similar
position of these galaxies in the diagnostic diagrams we suggest that
a similar interpretation might apply to IRAS\,23365. 

Finally we consider the population diagram for IRAS\,23420. 
Unfortunately we detect only lines from the $\nu$ = 1 level for this
galaxy which makes the interpretation inconclusive. Some of these
detections are also only weak, leading to uncertainties. Our model
finds 30 per cent of the 1--0S(1) flux, 70 per cent of the total H$_{2}$ flux,
having a non-thermal origin, with a temperature of 2450K, agreeing
with the conclusions drawn previously using the line ratios.

\subsection{Correlation between H$_{2}$ $\nu$ = 1-0S(1) and
[OI]$\lambda$6300} 

\begin{figure}
\centerline{\psfig{figure=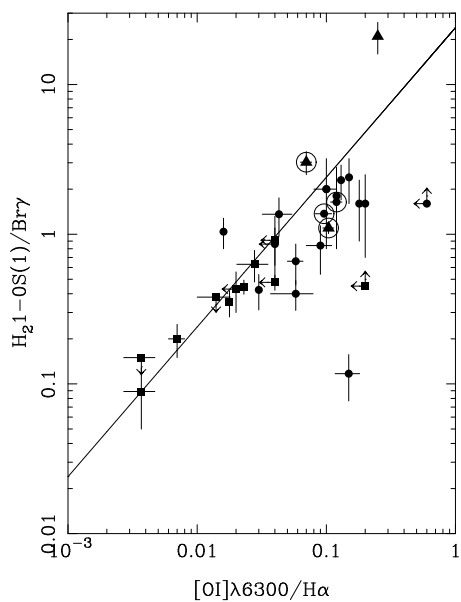,width=6cm}}
\caption{A plot of H$_{2}$ $\nu$ = 1--0S(1)/Br$\gamma$ and
[OI]$\lambda$6300/H$\alpha$. Symbols are as follows: squares
represent starburst nuclei, circles represent AGN and triangles
represent LINERS. Circled points are from this work, all others from
Mouri et al. \shortcite{mh89}. Solid line represents best fit to
starburst nuclei data. 
\label{o1_plot}}
\end{figure}

Mouri et al. \shortcite{mh89} find a linear correlation between the
H$_{2}$ $\nu$=1--0S(1)/Br$\gamma$ and [OI]$\lambda$6300/H$\alpha$
emission line ratios for a sample of galaxies containing an AGN or a
starburst nucleus. Their data show both the H$_{2}$1-0S(1) (hereafter
S(1)) and [OI] lines are enhanced in AGN. Using optical data, where
available, from the literature we have added the galaxies we have
observed to those in the Mouri sample, see
Figure~\ref{o1_plot}. Squares represent starburst nuclei, circles
represent AGN and triangles represent LINERS. The circled points are
the four galaxies from our sample: IRAS\,16487, IRAS\,17179,
IRAS\,20210 and IRAS\,23365. Classifications for these galaxies, taken
from the literature, are LINER, Sy2, Sy2 and LINER respectively. The
positions of these galaxies on this plot are as expected; they lie
amongst the AGN with enhanced S(1) and [OI] emission. 

Mouri et al. \shortcite{mh89} have only one LINER included in their
sample, NGC\,6240 (mentioned earlier), which they note has a
remarkable S(1) line width with FWHM $\sim$ 650 km s$^{-1}$. When
NGC\,6240 is plotted it falls at the extreme upper right, away from
the other galaxies. The authors interpreted this as a special case in
which the emission is excited by shocks directly, due to the merger of
two galaxies, in addition to the effects of a central engine and/or
starburst activities. This theory has been supported by recent work on
NGC\,6240 by Ohyama et al. \shortcite{oy00}, who have found evidence
for young superbubbles in the central region.  They propose that the
interaction of this expanding shell with the extragalactic molecular
gas produces shock-excited intense H$_{2}$ emission.

The LINER galaxies we have added however, IRAS\,16487 and IRAS\,23365,
show smaller S(1) line widths (respectively FWHM $\sim$ 400 km
s$^{-1}$ and $\sim$ 250 km s$^{-1}$), consistent with their position
amongst the AGN.

\subsection{[Fe II] emission}

The most prominent [FeII] transition lines seen in the NIR are
a$^{6}$D$_{9/2}$ -- a$^{4}$D$_{7/2}$ at 1.257$\mu$m and
a$^{4}$F$_{9/2}$ -- a$^{4}$D$_{7/2}$ at 1.644$\mu$m. Due to the range of 
redshifts of the objects observed, the former line was not covered by
our data. The wavelength of the line at 1.644$\mu$m meant that it could
potentially be observed in every galaxy. Unfortunately, due the range
of redshifts observed, this line falls in the region where atmospheric
absorption is high for some objects. This is the case for IRAS\,00150,
IRAS\,17179 and IRAS\,23420. Despite this the [FeII]1.644$\mu$m line is
detected in all the galaxies observed except IRAS\,00150, where a
slight increase in flux can be seen at the appropriate
wavelength. Another, lower flux, [FeII] emission line at 1.9670$\mu$m
falls in the waveband observed. This line may be detected in IRAS\,17179
and IRAS\,20210. An increase in flux at the appropriate wavelength can
be seen for IRAS\,23365 and IRAS\,23420, however there is no detection
above the 2 $\sigma$ limit. At this resolution, this line would be
blended with [SiVI] emission. Such an apparently low incidence of the
[FeII] 1.9670$\mu$m line in this sample is perhaps slightly surprising
when compared with a recent survey of ULIRGs where detections of this
line are claimed for 81 per cent of objects observed \cite{mt00}. However
the typical [FeII]/Pa$\alpha$ ratio quoted is around 1 per cent for objects
with weaker [FeII] emission, rendering the line too faint to be seen
for the objects we observe. We note Murphy et al. \shortcite{mt00} do
not detect [FeII] 1.9670$\mu$m for IRAS\,16487 or IRAS\,23365. 

The excitation mechanism of [FeII] emission is still under
debate. Like [OI], [FeII] emission is excited by 
electron collisions in partially ionised regions (PIRs). 
Mouri et al. \shortcite{mh00} attribute extensive PIRs in starburst
galaxies to shock heating from supernovae, whereas in Seyfert
galaxies they propose these regions arise due to photoionisation by
nuclear radiation. Alonso-Herrero et al. \shortcite{aa97} demonstrate
a similar idea, favouring a progression of the proportion of shock
excitation present from HII regions to starburst galaxies, through to
Seyferts and supernovae. This is demonstrated in the diagnostic diagram 
they employ where the line ratios of [FeII]1.644/Br$\gamma$ vs
[OI]6300/H$\alpha$ produce a separation of the galaxy types, indicating 
the proposed progression.  

When the objects observed here are added to the figure from
Alonso-Herrero et al. \shortcite{aa97} they fall amongst the other
galaxies, in the area marked by the authors as the overlap between
Seyferts and composite galaxies. From this we would infer a mixture of 
shock heating and photoionisation as the mechanism for the production
of [FeII] in these galaxies, with the latter process appearing to
dominate. Overall, based on these ratios the emission seems to be
typically Seyfert-like, as expected for these galaxies.

\subsection{[SiVI] emission}
Detection of [SiVI] emission indicates the presence of high energy
ionising photons. This high excitation line has an ionisation
potential of 167\,eV and cannot be formed by continuum emission from
young stars. It is a feature only associated with active galaxies,
suggesting that the mechanism for its production is photoionisation by
an active nucleus \cite{ma94}. The [SiVI] line is detected for
IRAS\,17179 and 
the main nucleus of IRAS\,20210, having FWHM $\sim$~1200 km s$^{-1}$
in both detections. This detection confirms the status of these two
galaxies as Seyferts. [SiVI] is not detected in the spectra of the
LINER type galaxies. The absence of this line in the spectra of the
galaxies of undefined type: IRAS\,00150 and IRAS\,23420, does not
support the presence of a Seyfert nucleus, however the presence of a
highly obscured AGN cannot be ruled out using this criteria alone
\cite{ma94}.  

\subsection{CO absorption}
\label{co}
\begin{figure}
\centerline{\psfig{figure=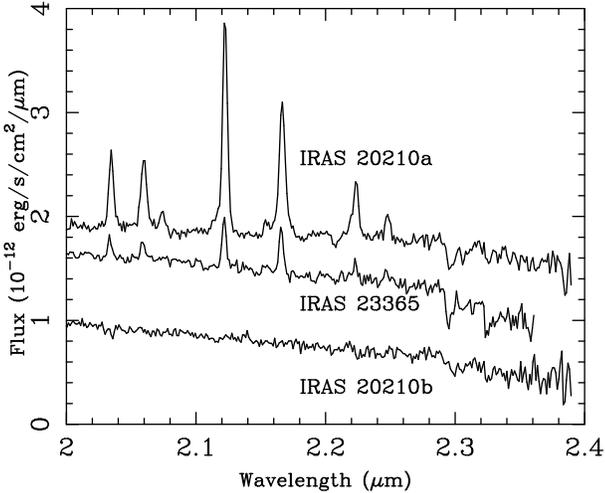,width=8cm}}
\caption{Section of K band spectra for IRAS\,20210 (both nuclei) and
IRAS\,23365 showing CO absorption in more detail. Note: objects are all
plotted 'as is' and have not been scaled for viewing purposes.
\label{coband}}
\end{figure}

CO absorption arises from the presence of older stars, red giants and
supergiants. There are three main bandheads potentially visible:
$^{12}$CO(2,0) at 2.294$\mu$m, $^{12}$CO(3,1) at 2.323$\mu$m and
$^{13}$CO(2,1) at 2.345$\mu$m. In our observations, due to the
redshift range of the objects observed, detection of these absorption
features was potentially only possible in two objects: IRAS\,20210 and
IRAS\,23365. Both objects show absorption, possibly visible in both
nuclei for IRAS\,20210. Figure~\ref{coband} shows an expanded view of
the spectra of the objects in the relevant waveband, useful to show
the CO absorption not obvious with the scaling in
Figure~\ref{spectra}. The equivalent width of the absorption in the
main nucleus of IRAS\,20210 is smaller than that of IRAS\,23365. This
is as expected, since emission from heated dust begins to become
significant around this wavelength, rising into the mid-infrared. In
IRAS\,20210, a Sy2, the stronger emission due to hot dust effectively
dilutes the measured CO absorption bands. In IRAS\,23365 the hot dust
continuum is less, allowing CO absorption to become more evident. In
the second nucleus of IRAS\,20210, labelled in Figure~\ref{coband} as
IRAS\,20210b, the emission is characteristic of an older stellar
population: showing CO absorption and a lack of recombination emission
lines. Additional evidence for the presence of a stronger 'diluting'
hot dust continuum comes from the relatively redder spectrum of
IRAS\,20210 when compared with IRAS\,23365.

\section{Conclusions}
\label{summary}

We have presented NIR observations of a sample of mainly
interacting ULIRGs, comprising of H and K band spectra. From the
analysis of these data our main aims were to investigate the power
source of these extremely luminous objects, the potential excitation 
mechanisms of the strong molecular hydrogen emission which is often
seen in such objects, along with an investigation of the other
characteristics such as extinction and multiple nuclei. Our results
are summarised below.

Of the seven galaxies observed, five have previous classifications and
the remaining two, IRAS\,00150 and IRAS\,23420, have no published
optical data other than that used for redshift determination.
Broadened HI recombination emission was detected in one object,
IRAS\,23498, with FWHM $\sim$ 5000 km s$^{-1}$. This finding is
consistent with previous results for this object which is classified
as a Sy2. The high excitation line [SiVI] is detected in two galaxies
in our sample, both are classified as Sy2s. We have investigated
several proposed correlations, which as well as exhibiting a specific
trend between line ratios, also help to distinguish AGN from starburst
galaxies. In most cases the galaxies added to the existing data
exhibited the expected trend. We question a correlation found by
Goldader et al. \shortcite{gj95} between L$_{Br\gamma}$ vs L$_{FIR}$
for LIRGs. This is due to the small size of the sample on which the
relationship is based. Data points added for the ULIRGs suggest either
(1) the need for a different form of correlation, or (2) that no
significant correlation is seen.

Regarding the classification of IRAS\,00150 and IRAS\,23420, neither
spectrum exhibits [SiVI] emission. However this does not rule out the
presence of highly obscured AGN. It would be useful to obtain good
quality optical data for these objects, in order to classify them.

Another objective was to provide information on the excitation
mechanism of the molecular hydrogen emission observed in the K
band. Due to the variation in the number of lines observed, several
diagnostic methods were used.

The overall finding for the galaxies in our sample, with the possible
exception of IRAS\,23365 and IRAS\,23420, is the predominance of
thermal excitation. Non-thermal emission is also present in some
objects but in varying proportions. Comparison of only a few line
ratios indicated very little, and possibly negligible, contribution
from non-thermal emission in IRAS\,00150, IRAS\,16487 and
IRAS\,17179. More detailed analysis was possible for the other objects
in the sample. IRAS\,20210 shows only a small proportion, 10 per cent,
of non-thermal contribution to the 1--0S(1) emission. In contrast,
values of 30 per cent non-thermal emission were found for both
IRAS\,23365 and IRAS\,23420. This corresponds to $\sim$70 per cent of
the total H$_{2}$ emission.

Since most of the galaxies observed are interacting, one might expect
thermal excitation through shocks to be the dominant mechanism. The
obvious exceptions to this are IRAS\,23365 (which indeed shows a
higher proportion of non-thermal emission), and also IRAS\,17179 and
IRAS\,23420 which are classified as non interacting galaxies.We note
that for IRAS\,17179 only a few H$_{2}$ emission lines were detected,
so our conclusions are less certain. However, the increased fraction
of non-thermal emission in IRAS\,23420 is consistent with it being
non-interacting. 

Values for reddening were derived for galaxies with available optical
data, using NIR emission lines in combination with the optical
lines. The results depended on the combination of lines used, with
values using only optical lines being consistently smaller than those
found using a NIR line with an optical line. This is as predicted
since the infrared lines can be detected from more highly reddened
regions.

Of the four galaxies with optical data, IRAS\,23365 showed the highest
values of extinction with an average value of A$_{V}$ = 4.7 $\pm$ 0.8
mag. In comparison the value found using optical lines only was
A$_{V}$ = 2.8 mag. The former value is consistent with that derived by
Davies et al. \shortcite{dr00b}, A$_{V}$ = 4.4 mag, found using a
model of 500K dust contributing 50 per cent of the K band emission
using the NIR imaging. The values of extinction found for the other
galaxies were moderate, being in the range A$_{V}$ = 2.3 -- 4.5 mag.

There is a bias in our sample towards interacting systems. For the two
for which we have spectra for both nuclei, we find either very weak or
no emission lines, indicative of an evolved population, in contrast to
the brighter primary nucleus. This trend has also been found for other
ULIRGs \cite{mt99,mt00}. An estimate of the contribution of the
secondary nucleus to the IR luminosity of the system was made and in
both cases the contribution was found to be minor. Observations of the
secondary nuclei of IRAS\,16487 and IRAS\,23498 would be interesting,
to see if these characteristics are shared by secondary nuclei of
interacting ULIRGs in general.

As mentioned above, all the hydrogen recombination lines detected were 
narrow, with the exception of IRAS\,23498. An interesting feature that
arises frequently in these objects is the blue shoulder of the
Pa$\alpha$ line. This feature could be explained by the presence of a
blue velocity component. Another possibility is the presence of
blended lines, as seen at our spectral resolution, higher
resolution spectroscopy is required to make further progress. 

\section{Acknowledgements}

AJB is supported by a PPARC studentship. The United Kingdom Infrared
Telescope is operated by the Joint Astronomy Centre on behalf of the
U.K. Particle Physics and Astronomy Research Council. This research
has made use of the NASA/IPAC Extragalactic Database (NED) which is
operated by the Jet Propulsion Laboratory, California Institute of
Technology, under contract with the National Aeronautics and Space
Administration. We would like to acknowledge an anonymous referee for
helpful comments and suggestions.

\appendix

\section{Details for individual galaxies}
\label{app}
A short synopsis is given here for each galaxy observed. It will
include a mention of previous work and a brief summary of the emission
lines seen.

\subsection{IRAS\,00150+4937}
Little previous data exists for this double nuclei galaxy which means
it is unclassified. It is amongst the faintest of the objects in the
sample with small EW for the lines detected. It is the only object in
the sample for which no blue wing of Pa$\alpha$ was seen. Detection of
only two H$_{2}$ emission lines made the determination of excitation
mechanism difficult. The ratios seen were indicative of shock
excitation over fluorescence. The spectra for the secondary nucleus of
this system showed very little, except weak Pa$\alpha$ emission.

\subsection{IRAS\,16487+5447}
This object was classified by Leech et al. \shortcite{lk94} as
isolated and undisturbed on the basis of photographic data. Further
NIR imaging \cite{cd96b,mt96} has shown this object is interacting,
although we only present spectra of the primary nucleus. The optical
classification of this object is LINER, while the ISO classification
is starburst \cite{ld99}. The Pa$\alpha$ line for this object shows a
blue wing. Only two H$_{2}$ emission lines were detected, from these
we conclude pure shock excitation is the most likely excitation
mechanism, consistent with the conclusions of Taniguchi et
al. \shortcite{ty99}.

\subsection{IRAS\,17179+5444}
This Seyfert 2 \cite{vs99a} was classified by Leech et
al. \shortcite{lk94} as being a interacting system. This work and
Davies et al. \shortcite{dr00b} show the secondary nucleus to be a
foreground star. This galaxy is one of the weaker of the sample and
its classification as a ULIRG depends on the value of H$_{o}$
used. Pa$\alpha$ shows a blue wing which could conceivably be broadened 
emission. [SiVI] is detected for this galaxy. Only two H$_{2}$
emission lines were detected, from these we conclude shock excitation
is the most likely excitation mechanism.

\subsection{IRAS\,20210+1121}
This interacting Seyfert 2 \cite{pe90,vj93} has been relatively well
studied in the past. The spectra we have obtained for this system are
the best of the sample with high S/N allowing identification of many
emission lines, included fainter H$_{2}$ excitation lines. Data in the
optical, NIR, radio and X-ray exist for this object. In this work this
object shows the broadest Pa$\alpha$ emission of the sample with FWHM
$\sim$ 700 km s$^{-1}$. This line also shows a blue wing which could
conceivably be further broadened emission. Analysis of the H$_{2}$
emission points towards a mainly thermal origin with a fraction of
fluorescence. [SiVI] emission is detected from this galaxy consistent
with its classification as a Sy2. Spectra for the secondary nucleus of 
this system were also obtained for this work. These spectra show no
emission lines. CO absorption bands were detected for both nuclei.

\subsection{IRAS\,23365+3604}
This single nucleus object is generally classified as a LINER
\cite{vs95} but has also been labelled as a composite galaxy
\cite{bw98}. It is one of the objects in this sample classified
as non-interacting but does show a disturbed morphology. Previous K
band spectra exist for this object \cite{gj95,vs99b}. Unfortunately
removal of atmospheric effects was not totally successful for this
object, making the spectrum noisy between $\sim$ 1.82$\mu$m and
1.95$\mu$m (observed wavelength). However this region was free of
expected emission due to the redshift of this object. Of the sample,
IRAS\,23365 showed some of the highest values for {\it E(B-V)} calculated
using NIR lines. The Pa$\alpha$ emission measured showed a blue wing
seen on some many of this objects. This line was not broadened,
consistent with previous work \cite{vs99b}. Enough H$_{2}$ lines were
detected to allow some analysis of the potential excitation
mechanism. Final conclusions attributed perhaps as much as $\sim$ 70 per cent 
of the total H$_{2}$ emission to non-thermal mechanisms. Fairly strong 
CO absorption bands are observed are observed in this object.

\subsection{IRAS\,23420+2227}
Little previous work and no classification exists for this relatively
faint galaxy which shows no sign of interaction. The spectra shown
here do not support the possibility that this galaxy containing an
AGN. The Pa$\alpha$ emission seen is narrow, possibly showing a blue
wing. Three H$_{2}$ lines are identified which indicate a mainly
thermal excitation mechanism.

\subsection{IRAS\,23498+2423}
This interacting Seyfert 2 \cite{vs99a} has the highest redshift  of 
our sample. We obtained only a H band spectrum for this object since K 
band spectra already exist \cite{vs97}. Consistent with this previous
work, our spectrum shows broadened Pa$\beta$ emission FWHM $\sim$ 5000 
km s$^{-1}$. 


\begin{thebibliography}{}

   \bibitem[\protect\citename{Alonso-Herrero et al.\ }1997]{aa97}
   Alonso-Herrero A., Rieke M. J., Rieke G. H., Ruiz M., 1997, ApJ 482,
   747   

   \bibitem[\protect\citename{Auriere et al.\ }1996]{am96} Auriere M.,
   Hecquet J., Coupinot G., Arthaud R., Mirabel I.\ F., 1996, A\&A,  312,
   387  

   \bibitem[\protect\citename{Baan et al.\ }1998]{bw98}
   Baan W.\ A., Salzer J.\ J., Lewinter R.\ D., 1998, ApJ,  509, 633  

   \bibitem[\protect\citename{Black \& van Dishoeck }1987]{bj87} Black
   J.\ H., van Dishoeck E.\ F., 1987, ApJ,  322, 412  

   \bibitem[\protect\citename{Cardelli et al.\ }1989]{cj89}
   Cardelli J. A., Clayton G. C., Mathis J. S., 1989, ApJ, 345, 245

   \bibitem[\protect\citename{Clements \& Baker\ }1996b]{cd96b}
   Clements D.\ L., Baker A.\ C., 1996b, A\&A,  314, 5

   \bibitem[\protect\citename{Clements et al.\ }1996a]{cd96}
   Clements D.L., Sutherland W.J., McMahon R.G., Saunders W., 1996a,
   MNRAS, 279, 477

   \bibitem[\protect\citename{Davies et al.\ }2000]{dr00a}
   Davies R., Ward  M., Sugai H., 2000, ApJ,  535, 735 

   \bibitem[\protect\citename{Davies et al.\ }2000b]{dr00b} Davies et
   al, 2000b, in preparation

   \bibitem[\protect\citename{DePoy} 1987]{dd87} DePoy D. L., 1987, Ph.D.\
   Thesis, Univ. of Hawaii  

   \bibitem[\protect\citename{Genzel et al.\ }1998]{gr98} Genzel, R.,
   Lutz, D.,  Sturm, E., Egami, E., Kunze, D., Moorwood, A.\ F.\ M.,
   Rigopoulou, D., Spoon, H.\ W.\ W., Sternberg, A., Tacconi-Garman,
   L.\ E., Tacconi, L., \&  Thatte, N.\ , 1998, ApJ, 498, 579  

   \bibitem[\protect\citename{Goldader et al.\ }1995]{gj95}
   Goldader J.D., Joseph R.D., Doyon R., Sanders D.B., 1995, ApJ, 444, 
   97
 
   \bibitem[\protect\citename{Harwit et al.\ }1987]{hw87} Harwit, M.,
   Houck, J.\  R., Soifer, B.\ T., \& Palumbo, G.\ G.\ C.\ , 1987,
   ApJ, 315, 28  

   \bibitem[\protect\citename{Heckman, Armus, \& Miley \ }1990]{ht90}
   Heckman, T.\  M., Armus, L., \& Miley, G.\ K.\ , 1990, ApJS, 74, 833 

   \bibitem[\protect\citename{Helou et al.\ }1985]{hg85} Helou  G.,
   Soifer B.\ T., Rowan-Robinson M., 1985, ApJL,  298, L7  

   \bibitem[\protect\citename{Jog \& Solomon \ }1992]{jc92} Jog, C.\
   J.\ \&  Solomon, P.\  M.\ , 1992, ApJ, 387, 152  

   \bibitem[\protect\citename{Kim }1995]{kd95} Kim D. -C.\ , 1995, Ph.D.\
   Thesis, Univ. of Hawaii  

   \bibitem[\protect\citename{Kim \& Sanders \ }1998]{kd98ii} Kim
   D. -C.\ -.\ \&  Sanders D.\ B.\ 1998, ApJS, 119, 41  

   \bibitem[\protect\citename{Kim et al.\ }1998]{kd98} Kim D. -C.\ -.,
   Veilleux S., Sanders D.\ B., 1998, ApJ, 508, 627  

   \bibitem[\protect\citename{Lawrence et al.\ }1986]{la86} Lawrence,
   A., Walker, D., Rowan-Robinson, M., Leech, K.\ J., \& Penston, M.\
   V.\ , 1986, MNRAS, 219, 687 

   \bibitem[\protect\citename{Lawrence et al.\ }1989]{l89}
   Lawrence A., Rowan-Robinson M., Leech K., Jones D. H. P.,
   Wall J. V., 1989, MNRAS, 240, 329

   \bibitem[\protect\citename{Lawrence et al. \ }1999]{la99} Lawrence, 
   A.\ et al.\ , 1999, MNRAS, 308, 897 

   \bibitem[\protect\citename{Leech et al.\ }1994]{lk94}
   Leech K.J., Rowan-Robinson M., Lawerence A., Hughes J.D., 1994,
   MNRAS, 267, 253

   \bibitem[\protect\citename{Leitherer et al.\ }1999]{lc99} Leitherer C., 
   Schaerer D., Goldader J.\ D., et al., 1999, ApJS, 123, 3 

   \bibitem[\protect\citename{Lutz et al.\ }1999]{ld99} Lutz D., Veilleux S., 
   Genzel R., 1999, ApJL,  517, 13 

   \bibitem[\protect\citename{Marconi et al.\ }1994]{ma94} Marconi A.,
   Moorwood A. F. M., Salvati M., Oliva E., 1994, A\&A, 291, 18  

   \bibitem[\protect\citename{Mouri et al.\ }1989]{mh89} Mouri H.,
   Taniguchi Y., Kawara K., Nishida M., 1989, ApJL,  346, L73 

   \bibitem[\protect\citename{Mouri\ }1994]{mh94} Mouri H., 1994,
   ApJ, 427, 777  

   \bibitem[\protect\citename{Mouri et al.\ }2000]{mh00} Mouri H.,
   Kawara K., Taniguchi Y., 2000, ApJ, 528, 186  

   \bibitem[\protect\citename{Murphy et al.\ }1996]{mt96}
   Murphy T.W., Armus L., Matthews K., Soifer B.T., Mazzarella J.M.,
   Shupe D.L., Strauss M.A., Neugebauer G., 1996, ApJ, 111, 1025 

   \bibitem[\protect\citename{Murphy et al.\ }1999]{mt99} Murphy
   T. W., Soifer B. T., Matthews K., Kiger J. R., Armus L., 1999,
   ApJL, 525, 85   

   \bibitem[\protect\citename{Murphy et al.\ }2000]{mt00} Murphy
   T. W., Soifer B. T., Matthews K., Armus L., Kiger J. R., in press
   (astro-ph/0010077)

   \bibitem[\protect\citename{Ohyama et al.\ }2000]{oy00} Ohyama, Y.\ et al.\ ,
   2000, PASJ, 52, 563

   \bibitem[\protect\citename{Perez et al.\ }1990]{pe90} Perez E.,
   Manchado A.,Garcia-Lario P., Pottasch S.\ R., 1990, A\&A, 227, 407  

   \bibitem[\protect\citename{Puxley\ }1991]{pp91} Puxley P. J.,
   1991, MNRAS, 249, 11

   \bibitem[\protect\citename{Rigopoulou et al.\ }1996]{rig96}  
   Rigopoulou D., Lawrence A., White G. J., Rowan-Robinson M.,
   Church S. E., 1996, A\&A, 305, 747

   \bibitem[\protect\citename{Sanders et al.\ }1989]{sd89}
   Sanders D. B., Phinney E. S., Neugebauer G., Soifer B. T.,
   Matthews K., 1989, ApJ, 347, 29 

   \bibitem[\protect\citename{Sanders et al.\ }1988]{sd88}
   Sanders D. B., Soifer B. T., Elias J. H., Madore B. F.,
   Matthews K., Neugebauer G., Scoville N. Z., 1988, ApJ, 325, 74

   \bibitem[\protect\citename{Sanders and Mirabel\ }1996]{sm96} 
   Sanders D. B., Mirabel I. F., 1996, ARA\&A, 34, 725

   \bibitem[\protect\citename{Schlegel et al.\ }1998]{sd98}
   Schlegel D. J., Finkbeiner D. P., Davis M., 1998, ApJ, 500, 525 

   \bibitem[\protect\citename{Sugai et al.\ }1997]{sh97} Sugai H.,
   Malkan M.\ A., Ward M.\ J., Davies R.\ I., McLean I.\ S., 1997,
   ApJ, 481, 186  

   \bibitem[\protect\citename{Surace et al.\ }2000]{sj00} Surace J.\
   A., Sanders D.\ B., Evans A.\ S., 2000, ApJ,  529, 170  

   \bibitem[\protect\citename{Tanaka et al.\ }1989]{tm89} Tanaka
   M., Hasegawa T., Hayashi S. S., Brand P. W. J. L., Gatley
   I., 1989, ApJ, 336, 207  

   \bibitem[\protect\citename{Tanaka et al.\ }1991]{tm91} Tanaka
   M., Hasegawa T., Gatley I., 1991, ApJ, 374, 516   

   \bibitem[\protect\citename{Taniguchi et al.\ }1999]{ty99}
   Taniguchi, Y., Yoshino, A., Ohyama, Y., \& Nishiura, S.\ , 1999,
   ApJ, 514, 660  

   \bibitem[\protect\citename{Vader et al.\ }1993]{vj93} Vader J.\ P.,
   Frogel J.\  A., Terndrup D.\ M., Heisler C.\ A., 1993, AJ,  106, 1743 

   \bibitem[\protect\citename{Veilleux and Osterbrock\ }1987]{vo87}
   Veilleux S. \& Osterbrock, D.E., 1987, ApJS, 63, 295

   \bibitem[\protect\citename{Veilleux et al.\ }1995]{vs95} 
   Veilleux S., Kim D.-C., Sanders D.B., Mazzarella J.M., Soifer B.T,
   1995, ApJS, 98, 171 

   \bibitem[\protect\citename{Veilleux et al.\ }1997]{vs97} Veilleux
   S., Sanders D. B., Kim D.-C., 1997, ApJ, 484, 92 

   \bibitem[\protect\citename{Veilleux et al.\ }1999a]{vs99a} 
   Veilleux S., Kim D.-C., Sanders D.B., 1999a, ApJ, 522, 113

   \bibitem[\protect\citename{Veilleux et al.\ }1999b]{vs99b} 
   Veilleux S., Sanders D.B., Kim D.-C., 1999b, ApJ, 522, 139

   \bibitem[\protect\citename{Young et al.\ }1996]{ys96}
   Young S., Hough J. H., Efstathiou A., Wills B. J., Bailey
   J. A., Ward M. J., Axon D. J., 1996, MNRAS, 281, 1026 

   \bibitem[\protect\citename{Zhenlong et al.\ }1991]{zz91} 
   Zhenlong Z., Xiaoyang X., Zugan D., Hongjun S., 1991, MNRAS, 252,
   593 

\end{thebibliography}
\end{document}